\documentclass{PoS}

\usepackage{amsmath}
\usepackage{graphicx}
\usepackage{url}

\title{Temperature dependence of $\eta/s$: uncertainties from the equation of state}

\ShortTitle{Uncertainties of $(\eta/s)(T)$ from EoS}

\author{\speaker{Jussi Auvinen},$^a$,
        Kari J.~Eskola,$^{b,c}$
        Pasi Huovinen,$^d$
        Harri Niemi,$^{b,c}$
        Risto Paatelainen$^e$
        and P\'eter Petreczky$^f$\\
\llap{$^a$}Institute of Physics Belgrade\\
           Belgrade, Serbia\\
\llap{$^b$}University of Jyv\"askyl\"a\\
           Department of Physics\\
           P.O. Box 35, FI-40014 University of Jyv\"askyl\"a, Finland\\
\llap{$^c$}Helsinki Institute of Physics\\
           P.O. Box 64, FI-00014 University of Helsinki, Finland\\
\llap{$^d$}University of Wroc\l{}aw\\
           Wroc\l{}aw, Poland\\
\llap{$^e$}CERN\\
           Geneva, Switzerland\\
\llap{$^f$}Physics Department\\
           Brookhaven National Laboratory\\
           Upton, NY 11973. USA\\

E-mail: \email{auvinen@ipb.ac.rs}}

\abstract{We perform a global model-to-data comparison on Au+Au
  collisions at $\sqrt{s_{NN}}=200$ GeV and Pb+Pb collisions at $2.76$
  TeV and $5.02$ TeV, using a 2+1D hydrodynamics model with the EKRT
  initial state and a shear viscosity over entropy density ratio
  $(\eta/s)(T)$ with a linear $T$ dependence. To quantify the amount of
  uncertainty due to the choice of the equation of state (EoS), we compare
  analysis results based on four different EoSs: the well known
  $s95p$ parametrisation, an updated parametrisation based on the
  same list of particles in hadron resonance gas, but using recent
  lattice results for the partonic part of the EoS, and two new
  parametrisations based on the Particle Data Group 2016 particle list
  and the recent lattice results. We find that the choice of the EoS
  does affect the favoured minimum value of $\eta/s$, although within
  the confidence limits of the analysis. On the other hand, our
  analysis hardly constrains the temperature dependence of $\eta/s$,
  no matter the EoS.}

\FullConference{XIII Quark Confinement and the Hadron Spectrum - Confinement2018\\
		31 July - 6 August 2018\\
		Maynooth University, Ireland}

\begin{document}

\section{Introduction}

Recent advancements in multi-parameter model-to-data comparison have
provided notable constraints on the minimum value and temperature
dependence of the shear viscosity over entropy density ratio,
$\eta/s$, of the matter produced in the heavy-ion collisions at RHIC
and the LHC. The results of a Bayesian analysis with a flexible
initial state parametrisation~\cite{Bernhard:2016tnd,Bass:2017zyn}
agree with the temperature dependence of $\eta/s$ found in the earlier
study using the EKRT pQCD + saturation + hydrodynamics
model~\cite{Niemi:2015qia}.

However, with the exception of papers like
Refs.~\cite{Pratt:2015zsa,Moreland:2015dvc,Alba:2017hhe}, the equation
of state is taken as given in models used to extract the $\eta/s$
value from the data. Furthermore, in many studies in the literature,
EoS parametrisation $s95p$ \cite{Huovinen:2009yb} was used, a
parametrisation which is based on by now outdated lattice
data~\cite{Bazavov:2009zn}. To find out whether the works
employing the $s95p$ parametrisation are still relevant, and
whether the present uncertainties in the lattice results affect the
value of $\eta/s$ extracted from the data, we perform a Bayesian
statistics based analysis using four different parametrisations of
EoS. We compare the data obtained in $\sqrt{s_{NN}}=200$ GeV Au+Au
collisions~\cite{Adams:2004bi,Abelev:2008ab} and Pb+Pb collisions at
$2.76$ TeV~\cite{Aamodt:2010cz,Adam:2016izf} and $5.02$
TeV~\cite{Adam:2016izf,Adam:2015ptt} to the results of EKRT +
hydrodynamics calculations~\cite{Niemi:2015qia,Niemi:2015voa} with a linear
parametric form for $(\eta/s)(T)$. The resulting probability
distributions for the best-fit parameters indicate not only whether the
most probable parameter values depend on the EoS used, but also
whether the difference is larger than the overall uncertainty in the
fitting procedure.

The structure of this proceedings article is the following: we
describe the main features of the hydrodynamics model and the
equations of state in Section~\ref{sec:model}. A summary of the
statistical analysis methods is given in Section~\ref{sec:stats}. The
results are presented in Section~\ref{sec:results} and summary in
Section~\ref{sec:summary}.

\section{Hydrodynamical model and the equations of state}
\label{sec:model}

We use a viscous 2+1D hydrodynamical model~\cite{Niemi:2015qia}
with a linear parametrisation of the temperature dependence of $\eta/s$:
\begin{equation}
 (\eta/s)(T)=\begin{cases}
  S_{HG}(T_{min}-T) + (\eta/s)_{min}, &T < T_{min}\\
 S_{QGP}(T-T_{min}) + (\eta/s)_{min}, &T > T_{min},
           \end{cases}
\end{equation}
where the free parameters are the minimum value of shear
viscosity over entropy density ratio $(\eta/s)_{min}$, the location of
the minimum in temperature $T_{min}$, and the slopes below and
above $T_{min}$, denoted by $S_{HG}$ and $S_{QGP}$, respectively.

The initial energy density distribution is determined using the
event-averaged EKRT minijet local saturation model
\cite{Paatelainen:2012at,Paatelainen:2013eea}:
\begin{equation}
 e(\vec{r}_T,\tau_s(\vec{r}_T)) = \frac{K_{sat}}{\pi}[p_{sat}(\vec{r}_T,K_{sat})]^4,
\end{equation}
where $p_{sat}(\vec{r}_T,K_{sat})$ is the local saturation scale,
$\tau_s(\vec{r}_T) = 1/p_{sat}(\vec{r}_T,K_{sat})$ is the local
formation time, and the proportionality constant $K_{sat}$ is one of
the free parameters of our model.

We utilise four different parametrisations of EoS, which all combine
hadron resonance gas at lower temperatures with lattice QCD at high
temperatures. We prefer to not use the lattice data to parametrise the
EoS below $T=155$ MeV temperature to allow energy and momentum
conserving particlization without any non-physical discontinuities in
temperature and/or flow velocity. Our baseline is the well-known
$s95p$ parametrisation~\cite{Huovinen:2009yb} where hadron
resonance gas containing hadrons and resonances below $M<2$ GeV mass
from the 2005 PDG summary tables~\cite{Eidelman:2004wy} is connected
to parametrised hotQCD data from Ref.~\cite{Bazavov:2009zn}.

\begin{figure}[t]
 \centering   \hfill
 \includegraphics[width=6cm]{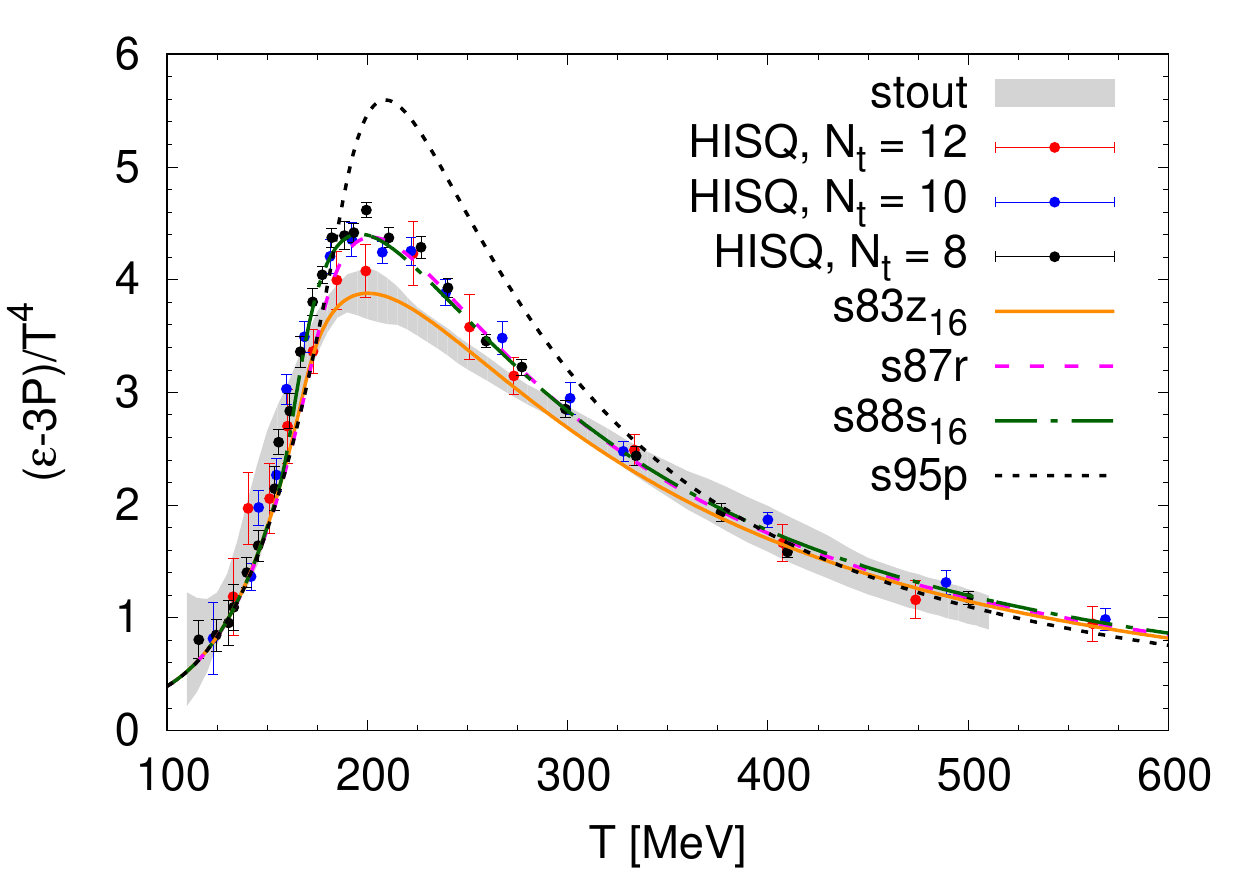}\hfill
 \includegraphics[width=6cm]{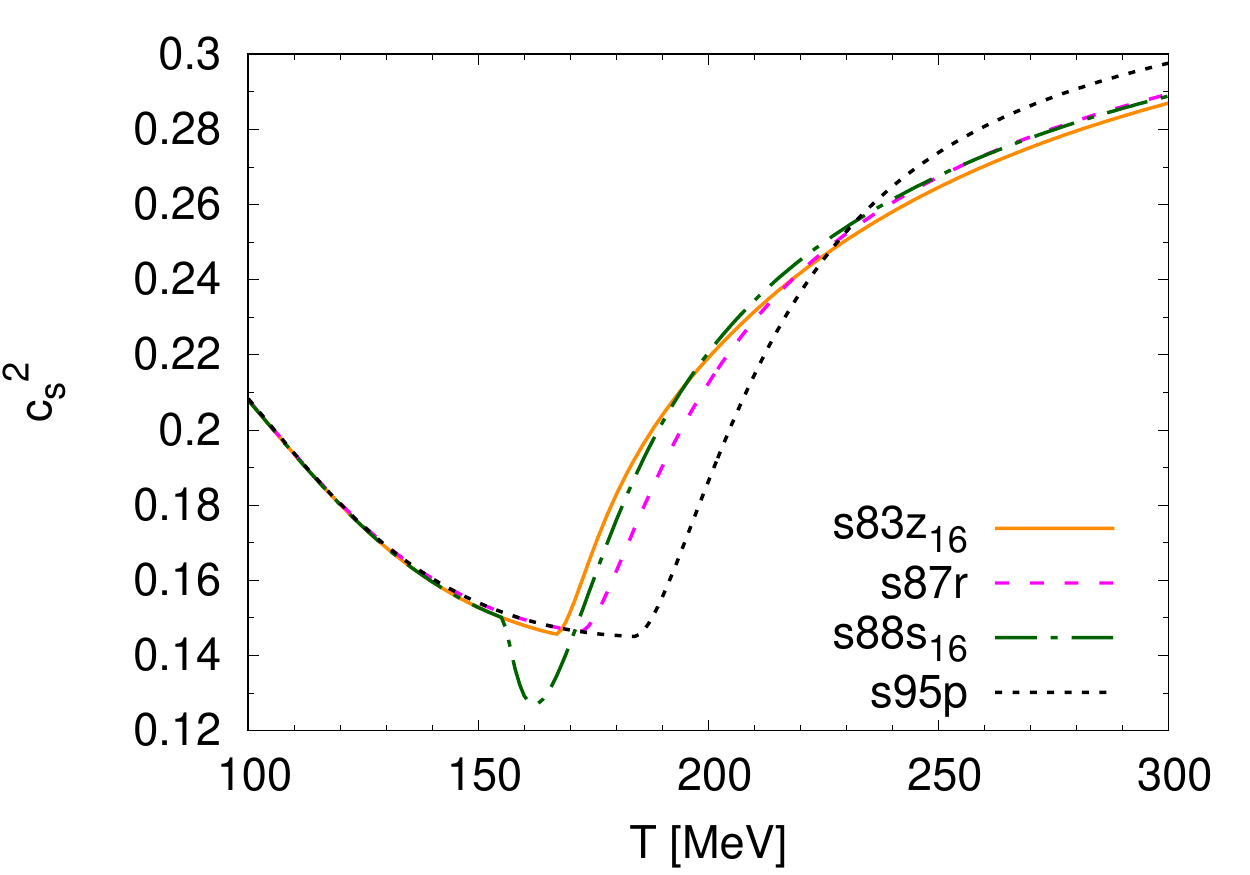}\hfill
 \caption{Left panel: The trace anomaly as a function of temperature in the four parametrisations of the EoS
   compared to the lattice data obtained using HISQ~\cite{Bazavov:2014pvz,Bazavov:2017dsy}
   and stout~\cite{Borsanyi:2013bia} discretisation schemes.  Right panel: The speed of sound squared
   as function of temperature in the four parametrisations of the EoS.}
 \label{fig:eos}
\end{figure}

To gauge the effect of various developments during the last decade, we
first connect the hadron resonance gas of PDG 2005 particle
list~\cite{Eidelman:2004wy} to parametrised 
set of contemporary lattice data
obtained using the HISQ discretisation scheme~\cite{Bazavov:2014pvz,Bazavov:2017dsy}.
To follow the convention used to
name $s95p$, we label this parametrisation $s87r$ since
entropy density reaches 87\% of its Stefan-Boltzmann value at $T=800$
MeV. The number of well-established resonances has increased since
2005, and thus we build our parametrisation $s88s_{16}$ based on
hadron gas containing all strange and non-strange hadrons and
resonances in PDG 2016 summary tables\footnote{Subscript ``16'' in the label of the
parametrisation refers to the use of the PDG 2016 list.}~\cite{Patrignani:2016xqp}, 
and on contemporary HISQ lattice data~\cite{Bazavov:2014pvz,Bazavov:2017dsy}.
Furthermore there is slight tension in the trace anomaly between the HISQ and
stout discretisation schemes, and to explore whether this
difference has any effect on hydrodynamical modeling, we base our
$s83z_{16}$ parametrisation on PDG 2016 resonances, and the continuum
extrapolated lattice data obtained using the stout discretisation~\cite{Borsanyi:2013bia}.
To characterise the differences in the parametrisations, we show the
trace anomaly and the speed of sound squared as a function of temperature in Fig.~\ref{fig:eos}.

In all the calculations presented here, the kinetic and chemical
freeze-out temperatures are $T_{\mathrm{dec}} = 150$ MeV, and bulk
viscosity is taken to be zero.

\section{Statistical analysis}
\label{sec:stats}

Our model has five free parameters
$\{K_{sat},T_{min},(\eta/s)_{min},S_{HG},S_{QGP}\}$ which need to be
fixed by looking for the best reproduction of experimental data.
Representing a point in the input parameter space with $\vec{x}$, a
corresponding point in the output (observable) space with
$\vec{y}(\vec{x})$, and the experimental data with
$\vec{y}^{\text{\,exp}}$, we can determine the posterior probability
distribution $P(\vec{x}|\vec{y}^{\text{\,exp}})$ of the best-fit
parameter values by utilising Bayes' theorem:
\begin{equation}
  P(\vec{x}|\vec{y}^{\text{\,exp}})\propto P(\vec{y}^{\text{\,exp}}|\vec{x})P(\vec{x}),
\end{equation}
where $P(\vec{x})$ is the prior probability distribution of input parameters
and $P(\vec{y}^{\text{\,exp}}|\vec{x})$ is the likelihood function
\begin{equation}
  P(\vec{y}^{\text{\,exp}}|\vec{x})
  = \frac{1}{\sqrt{|2\pi\Sigma|}}
     \exp\left(-\frac{1}{2}(\vec{y}(\vec{x})-\vec{y}^{\text{\,exp}})^T\Sigma^{-1}(\vec{y}(\vec{x})-\vec{y}^{\text{\,exp}}) \right),
\end{equation}
in which $\Sigma$ is the covariance matrix representing the uncertainties
related to the model-to-data comparison.

We produce samples of the posterior probability distribution with
Markov chain Monte Carlo (MCMC) \cite{ForemanMackey:2012ig}, by
initialising an ensemble of random walkers in the input parameter
space based on the prior probability, and accepting or rejecting each
proposed step based on the likelihood function. The distribution of
the taken steps then converges to the posterior distribution at a
large number of steps.

However, as we want to have an ensemble of $\mathcal{O}(100)$ walkers
performing $\mathcal{O}(1000)$ steps, it becomes infeasible to compute
the likelihood function by running the full hydrodynamics simulation
for each random input parameter combination $\vec{x}$. Instead, we
utilise Gaussian process (GP) emulators \cite{Rasmussen:2006} as a
surrogate model, and our likelihood function becomes
\begin{equation}
  P(\vec{y}^{\text{\,exp}}|\vec{x})
  = \frac{1}{\sqrt{|2\pi\Sigma|}}
     \exp\left(-\frac{1}{2}(\vec{y}_{GP}(\vec{x})-\vec{y}^{\text{\,exp}})^T\Sigma^{-1}(\vec{y}_{GP}(\vec{x})-\vec{y}^{\text{\,exp}}) \right),
\end{equation}
where $\vec{y}_{GP}(\vec{x})$ represents the GP estimate of the model output.
In this case, the covariance matrix includes also the emulator estimation error
\begin{equation}
\Sigma=\text{diag}((\sigma_{\text{\,exp}})^2+(\sigma_{\text{\,GP}}(\vec{x}))^2),
\end{equation}
with $\sigma_{\text{\,exp}}$ representing the experimental error
and $(\sigma_{\text{\,GP}}(\vec{x}))^2$ being the GP emulator variance.

\begin{figure}[h]
 \centering
 \includegraphics[width=6cm]{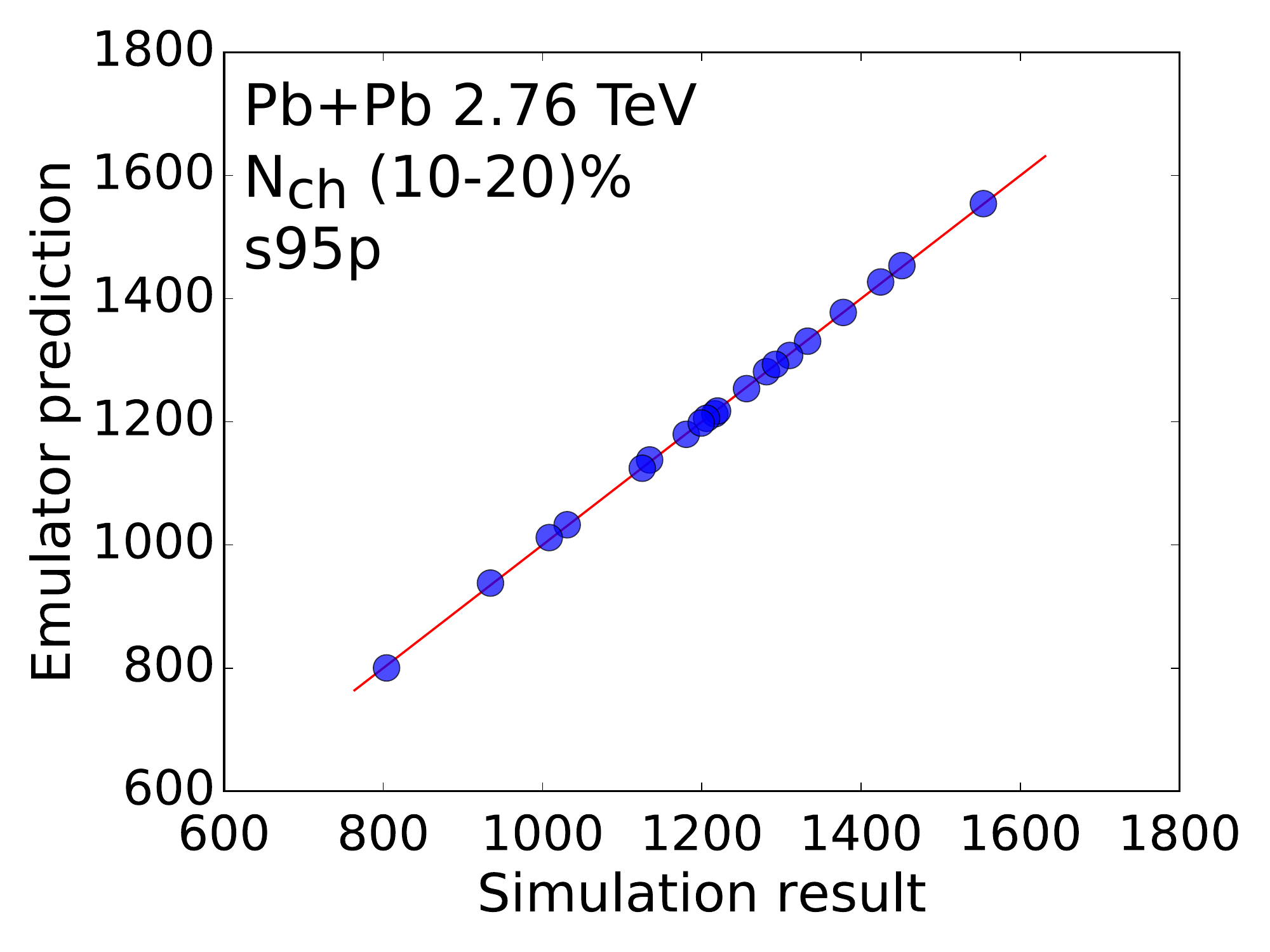}
 \includegraphics[width=6cm]{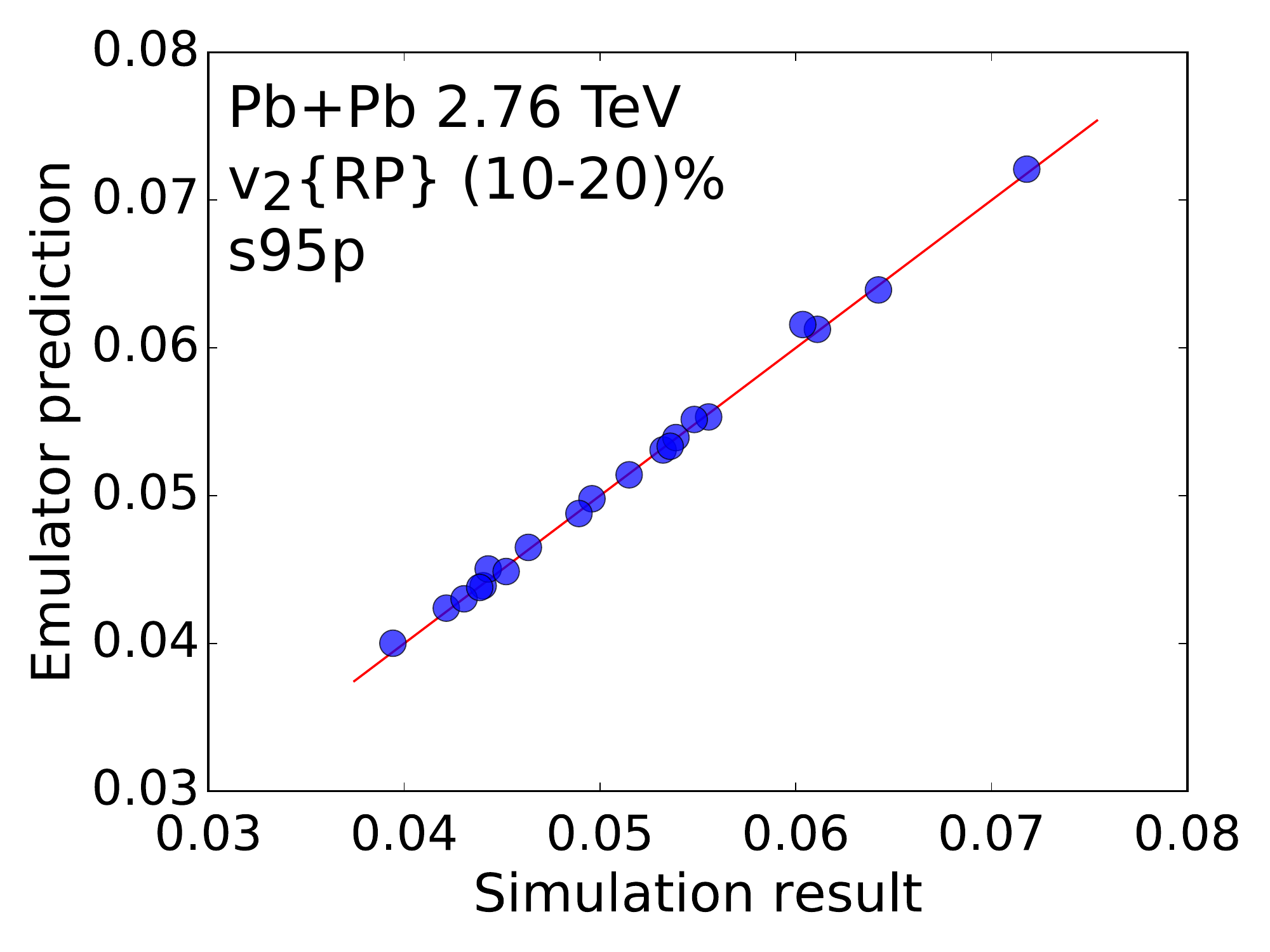}
 \caption{Illustration of the quality of the Gaussian process emulation for 20 random test points
 for simulations with $s95p$ equation of state.
 Left panel: Charged particle multiplicity in (10-20)\% most central Pb+Pb collisions at $\sqrt{s_{NN}}=2.76$ TeV.
 Right panel: Elliptic flow $v_2\{\text{RP}\}$ in (10-20)\% most central Pb+Pb collisions at $\sqrt{s_{NN}}=2.76$ TeV.}
 \label{fig:verification}
\end{figure}

The cost of using the GP emulators is paid in the form of running the
simulation multiple times with different parameter combinations to
create a set of training points for conditioning the GP. For the present
investigation, we have produced 120 training points for each EoS,
distributed evenly in the input parameter space using minmax Latin
hypercube sampling \cite{pyDOE:LHS}. The
emulation quality was then checked by removing 20 points from the
training set, conditioning the emulator on the remaining 100 and using
the emulator to predict the results at the 20 excluded points.
An example of the
results of this confirmation process is shown in
Fig.~\ref{fig:verification} for 2.76 TeV Pb+Pb collisions using
the $s95p$ parametrisation.

\section{Results}
 \label{sec:results}

In this preliminary study, the chosen small subset of the available
data to constrain the parameters consists of the charged particle multiplicity
$N_{\text{ch}}$ in $|\eta|<0.5$ and 4-particle cumulant elliptic flow
$v_2\{4\}$ in (10-20)\%, (20-30)\% and (30-40)\% centrality classes.
For Au+Au collisions at $\sqrt{s_{NN}}=200$ GeV, the $N_{\text{ch}}$
data was taken from Ref.~\cite{Abelev:2008ab} and $v_2\{4\}$ data from
Ref.~\cite{Adams:2004bi}.  For Pb+Pb collisions at
$\sqrt{s_{NN}}=2.76$ TeV, the charged particle multiplicities were
obtained from Ref.~\cite{Aamodt:2010cz} and for $\sqrt{s_{NN}}=5.02$
TeV from Ref.~\cite{Adam:2015ptt}.  Elliptic flow data for both
energies was taken from Ref.~\cite{Adam:2016izf}.

\begin{figure}[h]
 \centering
 \includegraphics[width=15cm]{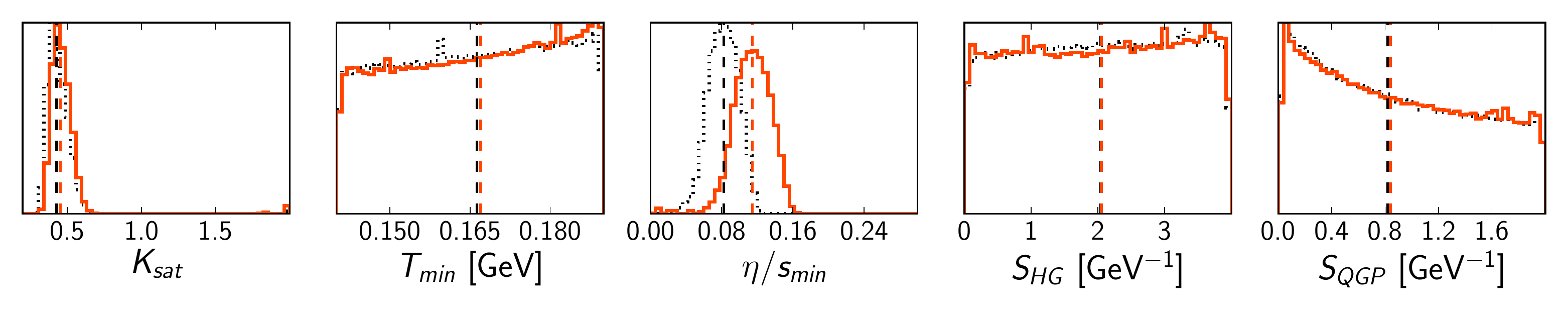}
 \caption{Marginal posterior probability distributions for EoS
   $s83z_{16}$ (orange solid lines) and $s95p$ (black dotted lines).
   Vertical dashed lines indicate the median values.
   The scale of the vertical axis on each panel is arbitrary.}
 \label{fig:posterior}
\end{figure}

Figure \ref{fig:posterior} shows the marginal posterior distributions
for each parameter (obtained from the full 5-dimensional probability
distribution by integrating over the other four parameters) for
the $s83z_{16}$ and $s95p$ equations of state. Our analysis is able to
find constraints for only two parameters, $K_{sat}$ and
$(\eta/s)_{min}$. For all EoSs $K_{sat}$ peaks at 0.5, which is in
agreement with the value used together with the {\em param1}
parametrisation in Ref.~\cite{Niemi:2015qia}, which in that study was
found to give the best agreement with the flow coefficients.  On the
other hand, the peak and thus the favoured value of $(\eta/s)_{min}$
does depend on the EoS.  The probability distributions of all the
other parameters, $T_{min},S_{HG},$ and $S_{QGP}$, are very flat
without clear peak, which indicates that our priors, \emph{i.e.} the
intervals where parameters were allowed to vary, were too narrow and
thus the median values of these distributions reflect mostly our
prejudices.

\begin{figure}[h]
 \centering
 \includegraphics[width=6cm]{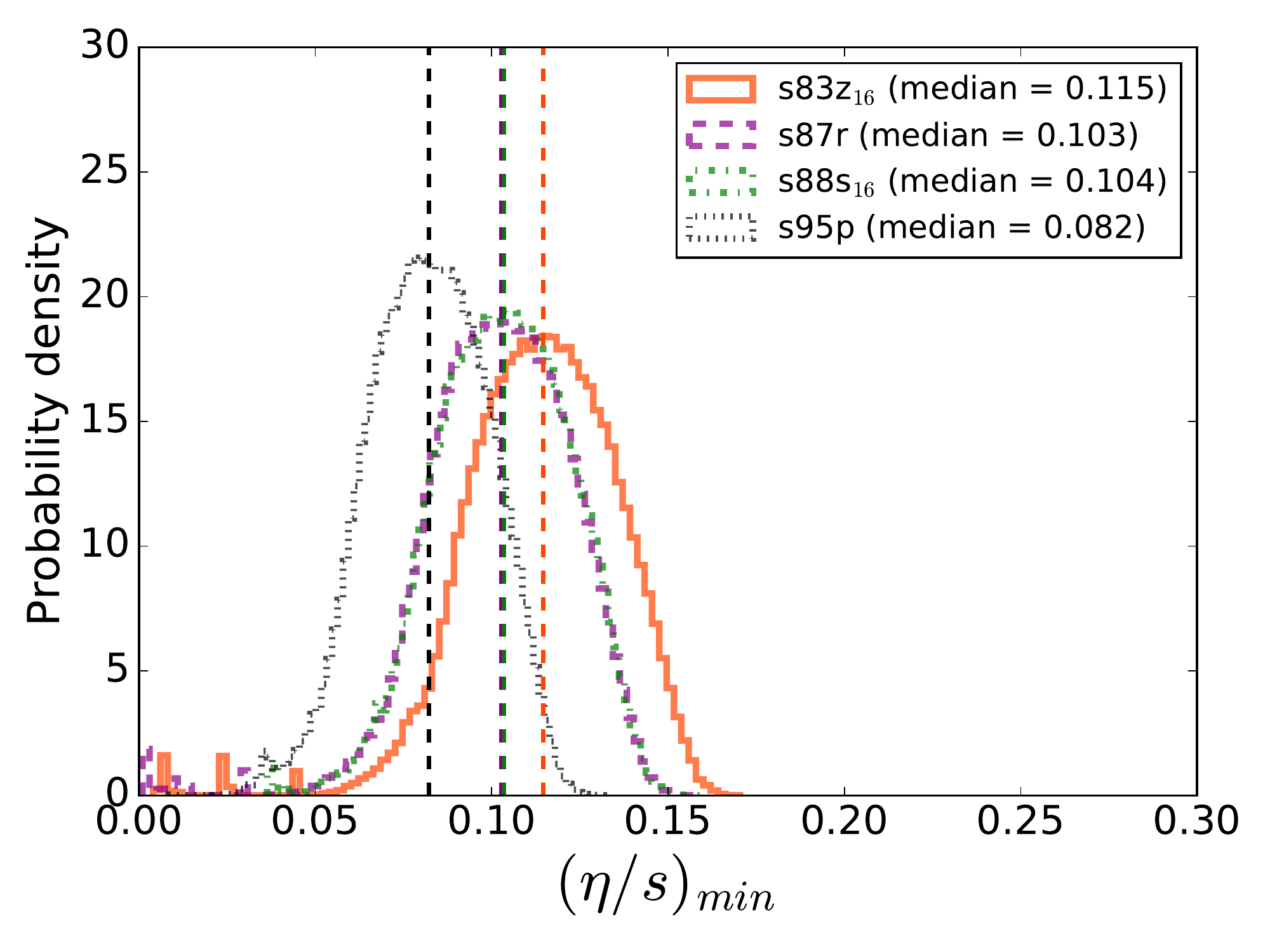}
 \includegraphics[width=6cm]{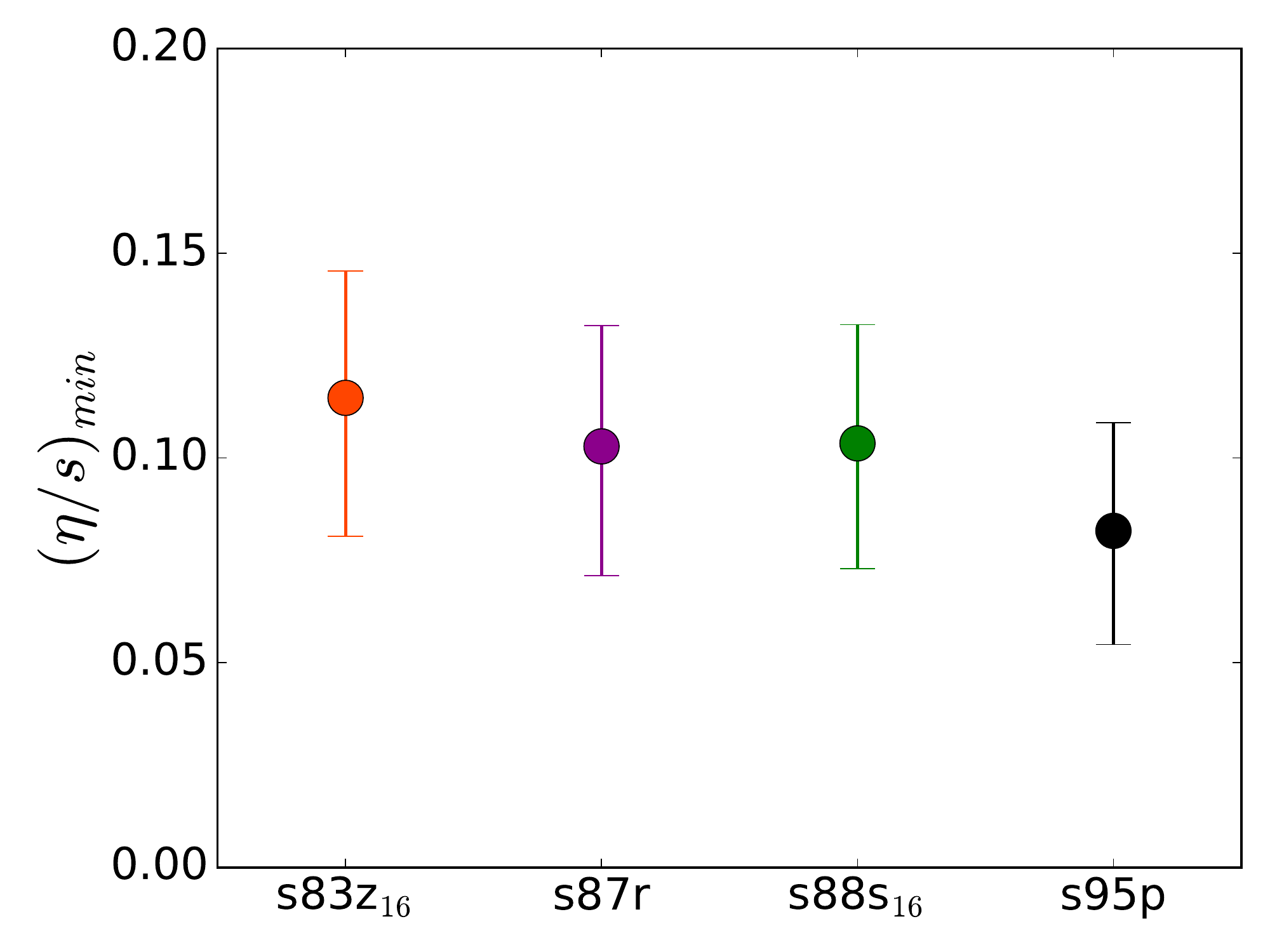}
 \caption{Comparison of $(\eta/s)_{min}$ posterior distributions for
   the four investigated EoSs. Left panel: Marginal probability
   distributions.  Dashed vertical lines indicate median values.
   Right panel: Simplified view showing the median values (filled
   circles) and 90\% credible intervals (error bars).}
 \label{fig:etasmin_comparison}
\end{figure}

We present a more detailed comparison of the most probable
$(\eta/s)_{min}$ values for all four EoSs in
Fig.~\ref{fig:etasmin_comparison}. The largest difference is observed
between $s95p$ and $s83z_{16}$, while the two other
parametrisations $s87r$ and $s88s_{16}$ fall in-between
with practically identical probability distributions. However, while
the median of the distribution is sensitive to the equation of state,
the 90\% credibility intervals overlap even for $s83z_{16}$ and
$s95p$ parametrisations. Therefore the earlier results obtained
using $s95p$ parametrisation are still valid within the overall
uncertainties of the fitting procedure, and it is too early to exclude
the possibility that all EoSs would produce similar values for
$(\eta/s)_{min}$ in a more detailed analysis.

\begin{figure}[h]
 \centering
 \includegraphics[width=4.9cm]{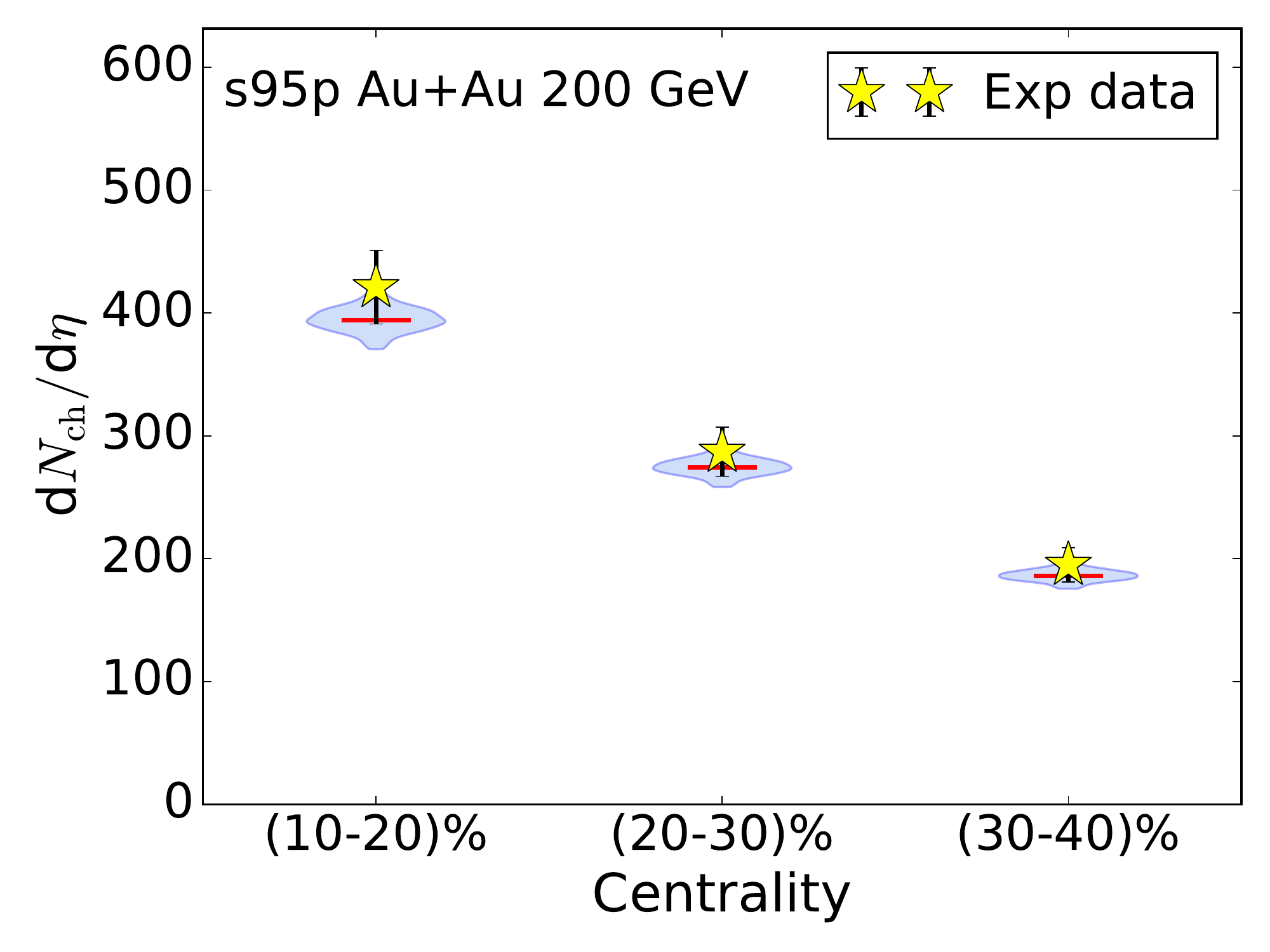}
 \includegraphics[width=4.9cm]{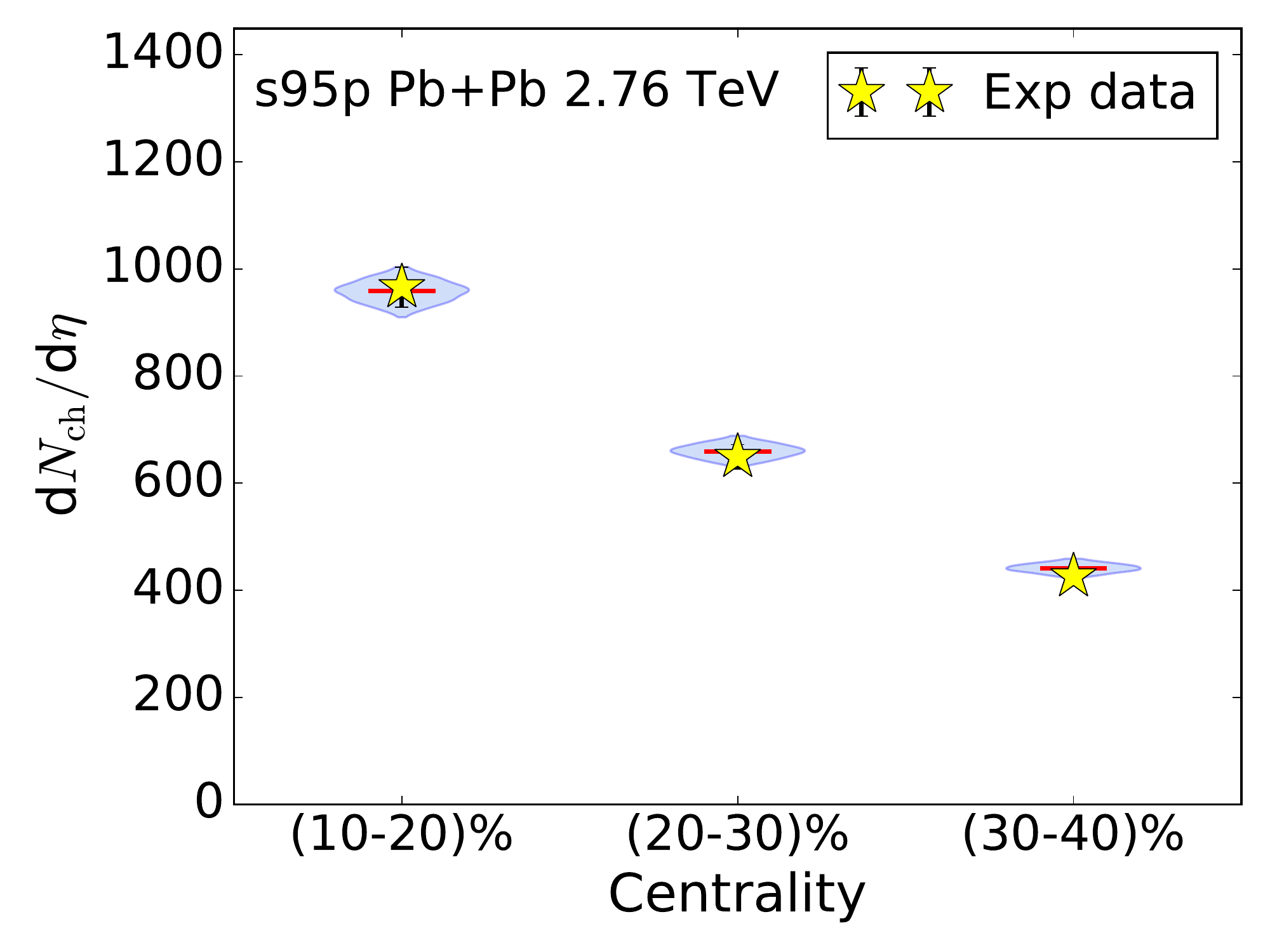}
 \includegraphics[width=4.9cm]{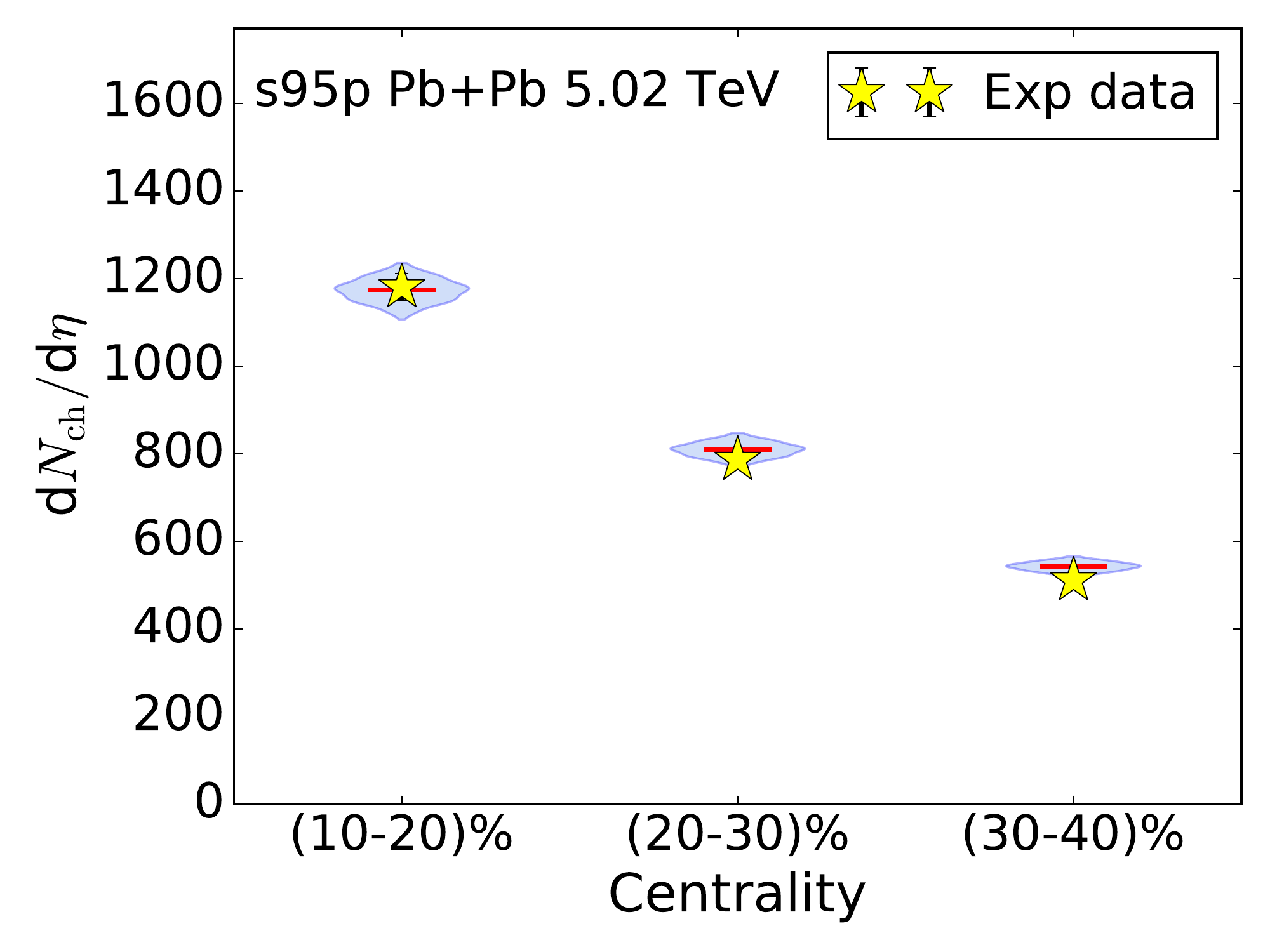}
 \caption{Predicted variation on charged particle multiplicity vs. centrality
 using 200 samples from the posterior distribution with the $s95p$ EoS.
 Wider shaded area indicates higher density of points.
 Horizontal lines within the shaded areas indicate the median values.
 Left panel: Au+Au at $\sqrt{s_{NN}}=200$ GeV compared to STAR data \cite{Abelev:2008ab}.
 Middle panel: Pb+Pb at $\sqrt{s_{NN}}=2.76$ TeV compared to ALICE data \cite{Aamodt:2010cz}.
 Right panel: Pb+Pb at $\sqrt{s_{NN}}=5.02$ TeV compared to ALICE data \cite{Adam:2015ptt}.}
 \label{fig:nch_posterior}
\end{figure}

\begin{figure}[h]
 \centering
 \includegraphics[width=4.9cm]{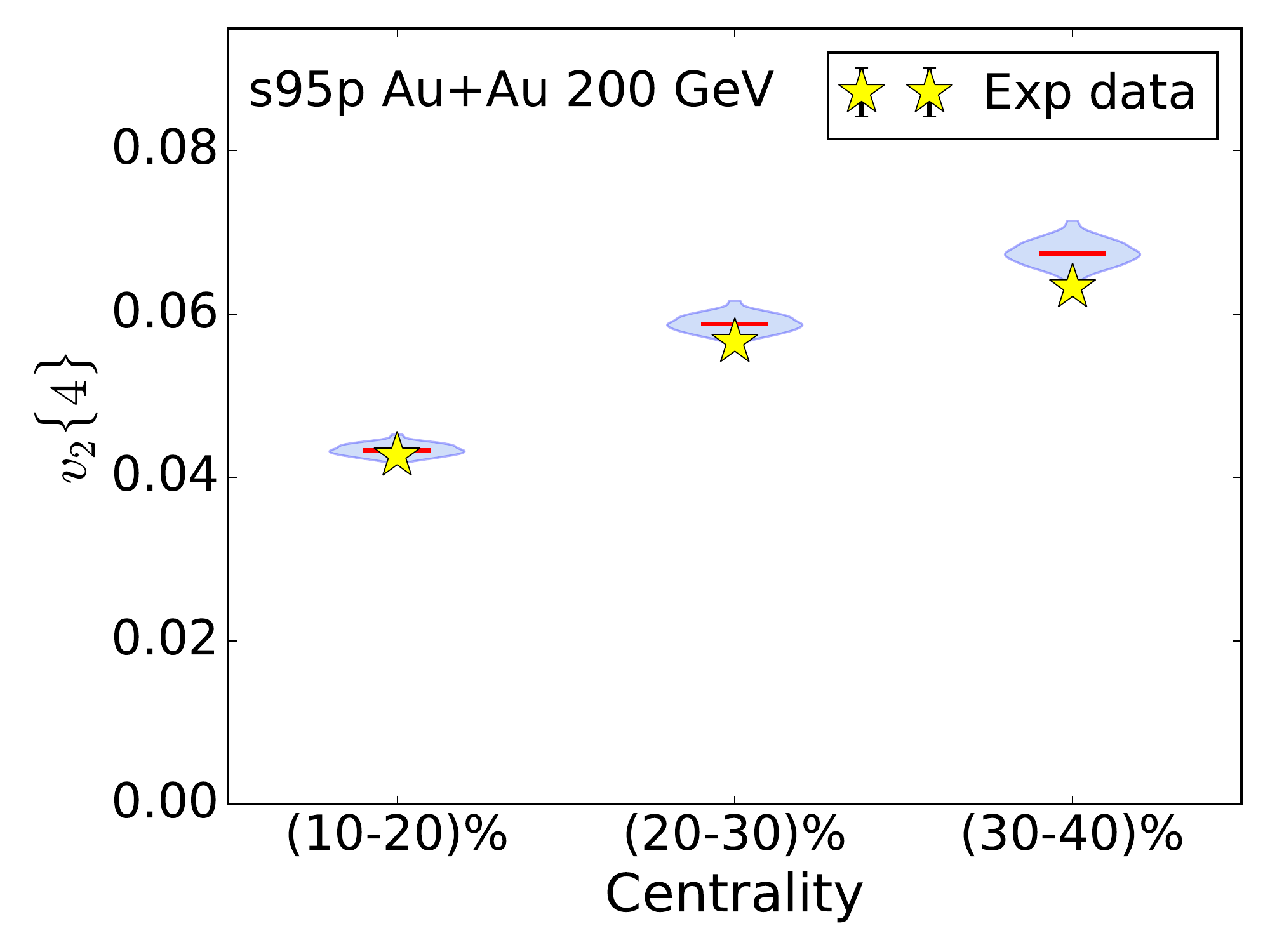}
 \includegraphics[width=4.9cm]{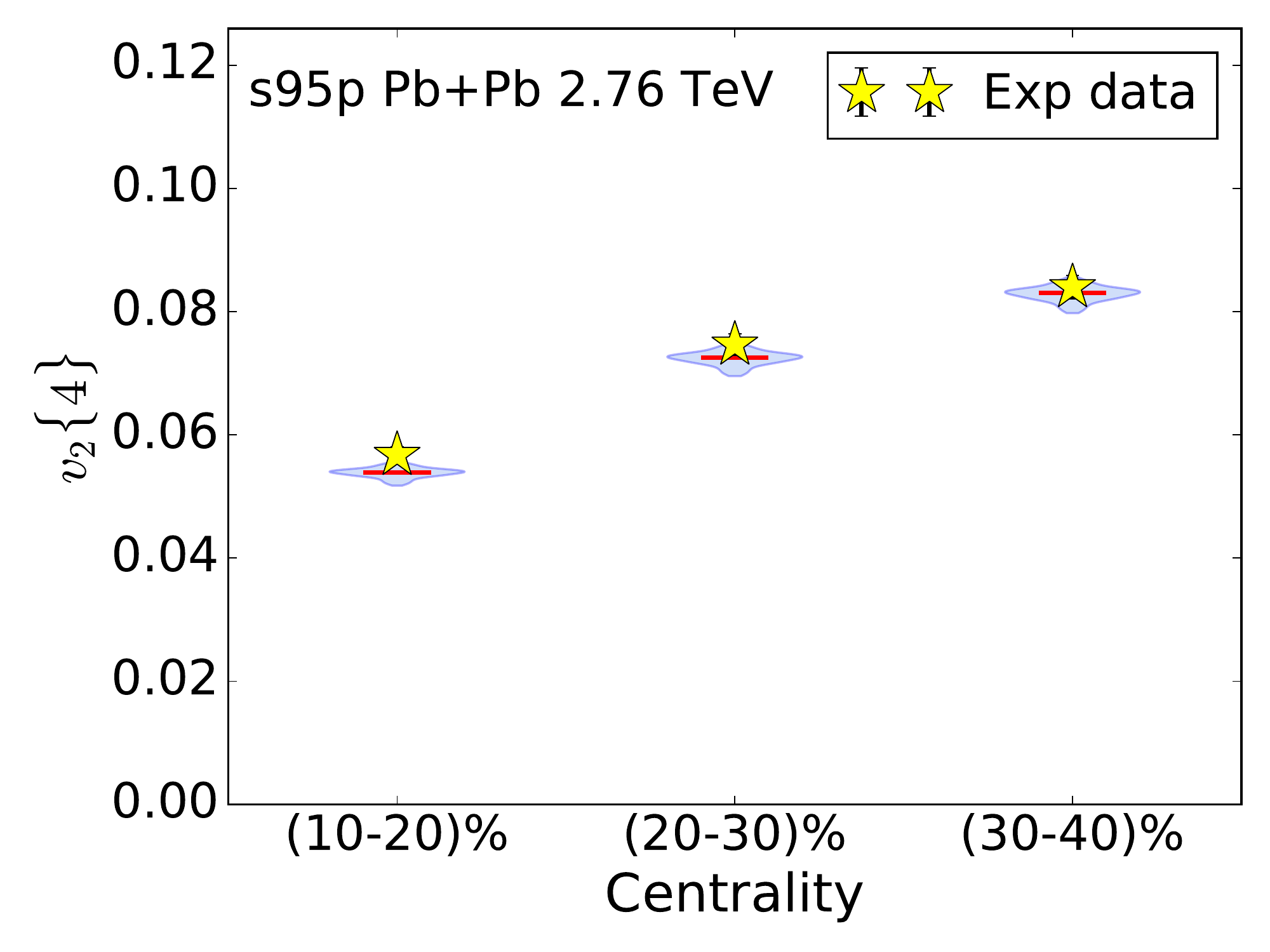}
 \includegraphics[width=4.9cm]{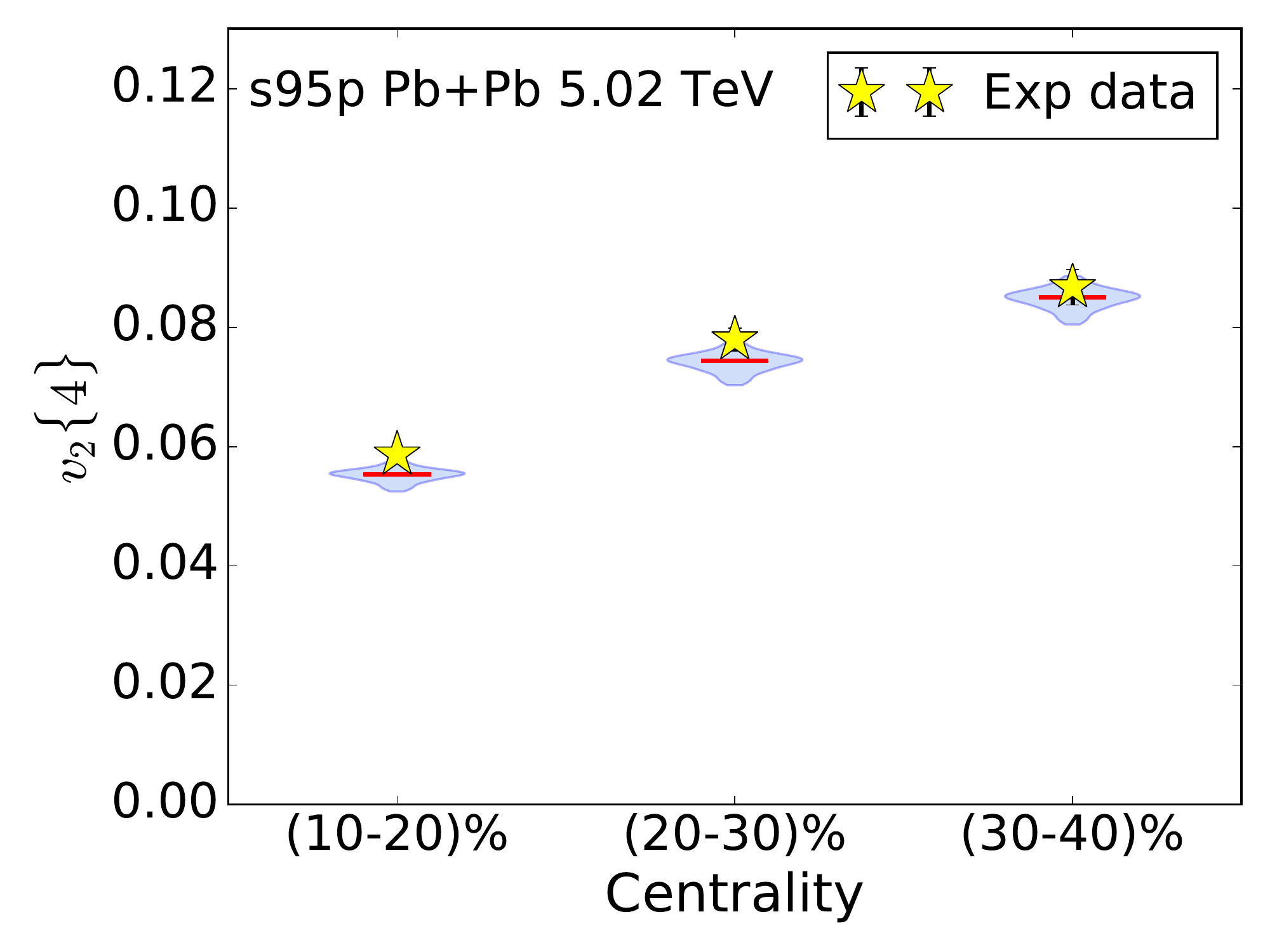}
 \caption{Predicted variation on elliptic flow vs. centrality
 using 200 samples from the posterior distribution with the $s95p$ EoS.
 Wider shaded area indicates higher density of points.
 Horizontal lines within the shaded areas indicate the median values.
 Left panel: Au+Au at $\sqrt{s_{NN}}=200$ GeV compared to STAR data \cite{Adams:2004bi}.
 Middle panel: Pb+Pb at $\sqrt{s_{NN}}=2.76$ TeV compared to ALICE data \cite{Adam:2016izf}.
 Right panel: Pb+Pb at $\sqrt{s_{NN}}=5.02$ TeV compared to ALICE data \cite{Adam:2016izf}.}
 \label{fig:v2_posterior}
\end{figure}

Finally, we check how well the favoured parameter combinations
reproduce the experimental data by drawing 200 samples from the
posterior distribution and using the Gaussian process emulator to
predict the simulation output for these values. The results for
charged particle multiplicities $N_{\text{ch}}$ and for the elliptic
flow $v_2\{4\}$ are shown in Figs.~\ref{fig:nch_posterior}
and~\ref{fig:v2_posterior}, respectively. The overall agreement with
the data is very good for both observables over all three collision
energies and centralities, assuring that the posterior distributions
are indeed providing best-fit parameter values.

\section{Summary}
\label{sec:summary}

We have determined the probability distributions of the best-fit
parameter values for shear viscosity over entropy density ratio
$\eta/s$ with a linear temperature dependence within a pQCD +
saturation + hydrodynamics framework using Bayesian statistics
approach. Using charged particle multiplicities and elliptic flow at
three different collision energies as calibration data, we were able
to find constraints for the initial state proportionality constant
$K_{sat}$ and the minimum value of shear viscosity $(\eta/s)_{min}$.
While $K_{sat}$ is found to be $\approx 0.5$ regardless of the choice
of the equation of state, the peak of $(\eta/s)_{min}$ distribution
depends on the EoS, and the three new parametrisations
$s83z_{16}$, $s87r$, and $s88s_{16}$ prefer larger
values of $(\eta/s)_{min}$ compared to the baseline EoS, $s95p$.
However, as the probability distributions still have a large overlap,
and the parameters controlling the temperature dependence of $\eta/s$
remain unconstrained, it is too early to make any strong statements
about the effect of EoS on the extracted shear viscosity.  In Bayesian
terms, better constraints are needed for both the prior (from theory)
and the likelihood (from measurements).

\section*{Acknowledgements}

JA was supported by the European Research Council, grant ERC-2016-COG:725741;
KJE and HN were supported by the Academy of Finland, Project no. 297058;
PH was supported by National Science Center, Poland, under grant Polonez
DEC-2015/19/P/ST2/03333 receiving funding from the European Union's
Horizon 2020 research and innovation programme under the Marie
Sk\l odowska-Curie grant agreement No 665778;
PP was supported by U.S.~Department of Energy under Contract No.~DE-SC0012704.

\bibliographystyle{JHEP}
\bibliography{auvinen_qconf}

\end{document}